\documentclass[aps,twocolumn]{revtex4}
\usepackage{graphicx}
\usepackage{bm}
\usepackage{dcolumn}% Align table columns on decimal point

\begin{document}

\title{Halfvortices in flat nanomagnets}

\author{Gia-Wei Chern}
\altaffiliation{Now at Department of Physics, University of
Wisconsin, Madison, Wisconsin 53706, USA}

\author{David Clarke}
\altaffiliation{Now at Department of Physics and Astronomy,
University of California, Riverside, California 92521, USA}

\author{Hyun Youk}
\altaffiliation{Now at Department of Physics, Massachusetts
Institute of Technology, Cambridge, Massachusetts 02139, USA}

\author{Oleg Tchernyshyov}
\affiliation{Department of Physics and Astronomy, The Johns Hopkins
University, Baltimore, Maryland, 21218, USA}

%\date{\today}

\begin{abstract}
We discuss a new type of topological defect in XY systems where the
O(2) symmetry is broken in the presence of a boundary. Of particular
interest is the appearance of such defects in nanomagnets with a
planar geometry. They are manifested as kinks of magnetization along
the edge and can be viewed as halfvortices with winding numbers $\pm
1/2$. We argue that halfvortices play a role equally important to
that of ordinary vortices in the statics and dynamics of flat
nanomagnets. Domain walls found in experiments and numerical
simulations are composite objects containing two or more of these
elementary defects. We also discuss a closely related system: the
two-dimensional smectic liquid crystal films with planar boundary
condition.
\end{abstract}

\maketitle

\section{Introduction}

It is well known that topological defects play an important role in
catalyzing the transitions between different ordered states for
systems with spontaneously broken symmetries
\cite{Chaikin,Mermin79}. For instance, in nanorings made of soft
ferromagnetic material, the switching process usually involves
creation, propagation, and annihilation of domain walls with complex
internal structure \cite{Zhu00,Klaeui03}. We have pointed out in a
series of papers\,\cite{Tch05,Chern06,Youk06} that domain walls in
nanomagnets of planar geometry are composed of two or more
elementary defects including ordinary vortices in the bulk and
fractional vortices confined to the edge. The simplest domain wall
in a magnetic strip consists of two edge defects with opposite
winding numbers $n=\pm 1/2$.

In a nanomagnet with the geometry of a disk, the strong shape
anisotropy due to dipolar interaction forces the magnetization
vector ${\bf M}$ to lie in the disk plane, effectively making the
magnet a 2D XY system. At the edge of the film, dipolar interaction
further aligns the spins to either of the two tangential directions
of the edge: $\hat{\bf m}={\bf M}/|{\bf M}| =\pm\hat{\bm{\tau}}$.
The reduction of ground-state symmetry from O(2) to a discrete $Z_2$
allows for a new type of topological defect confined to the edge.
These edge defects are manifested as kinks in magnetization
$\hat{\bf m}$ along the boundary. Kinks are topological defects
connecting different types of degenerate ground states in 1D systems
with discrete symmetries such Ising ferromagnet; their topological
properties usually are rather simple \cite{Chaikin}. Nevertheless,
as two of us pointed out in Ref.~\onlinecite{Tch05}, the edge
defects can be viewed as halfvortices and have nontrivial
topological charge related to the winding number of vortices in the
bulk.

\begin{figure}[t]
\includegraphics[width=0.88\columnwidth]{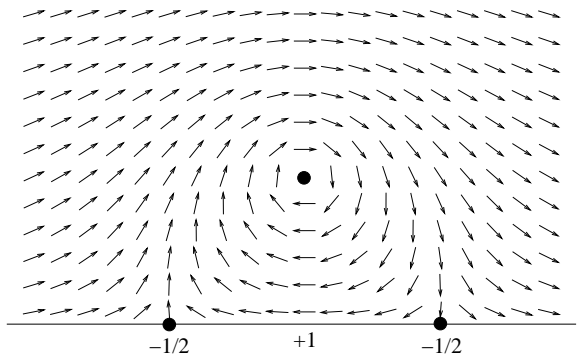}
\caption{A vortex ($n=+1$) absorbed by the edge can be viewed as its
annihilation with two $-\frac{1}{2}$ edge defects. The annihilation
results in a uniform magnetization pointing to the right.}
\label{fig-annihilation}
\end{figure}

For a bounded flat nanomagnet, the winding number of vortices in the
bulk is not a conserved quantity. This is illustrated by an example
shown in Fig.~\ref{fig-annihilation}, where a bulk vortex with
winding number $n=+1$ is absorbed into the edge. Conservation of
topological charges can be restored by assigning winding numbers to
edge defects. In this case there are two such kinks at the edge of
the film. The process shown in Fig.~\ref{fig-annihilation} then
expresses the annihilation of a $+1$ bulk vortex with two
$-\frac{1}{2}$ edge defects. Numerical simulations exhibiting
similar annihilation of bulk vortex with edge defects can be found
in Ref.~\onlinecite{Tch05}.

The winding number of a single edge defect is defined as the line
integral along the boundary $\partial\Omega$: \cite{Tch05}
\begin{equation}
    n = - \frac{1}{2\pi} \int_{\partial\Omega}
    {\bm \nabla}(\theta-\theta_\tau) \cdot d \mathbf{r}
    =\pm\frac{1}{2}.
\label{eq-n-e}
\end{equation}
Examples of edge defects with half-integer winding numbers are shown
in Fig.~\ref{fig-halfvortex-ex}. For a closed boundary the sum of
the winding numbers of edge defects is also given by the above
integral, but instead of integrating around one edge defect, the
integral is carried out along the entire boundary. It was shown in
Ref.~\onlinecite{Tch05} that this integral is related to the sum of
winding numbers of vortices in the bulk. In general, for a film with
$g$ holes, we obtain
\begin{equation}
\sum_{i}^{\rm edge} n_i + \sum_{i}^{\rm bulk} n_i = 1-g.
\label{eq-conserv}
\end{equation}
Here the winding numbers $n_i$ are integers for bulk defects and
half-integers for edge defects. This conservation law has important
implications for the dynamics of magnetization in nanomagnets
\cite{Tch05}. Since defects with large winding numbers carry
significant magnetic charge and are unfavored energetically in flat
nanomagnets, most of the intricate textures observed involve only
bulk vortices with winding number $n=\pm1$ and edge defects with
$n=\pm\frac{1}{2}$.

Topological considerations also place important constraints on the
possible structure of the domain walls in magnetic nanostrips
\cite{Chern06}. Examples of such domain walls are shown in
Figs.~\ref{fig-trans-wall} and~\ref{fig-vortex-wall}. Since edge
defects are kinks of magnetization along the boundary, a domain wall
in a magnetic strip must contain an odd number of kinks at each
edge. Furthermore, the angle of magnetization rotation along the two
edges must be compensated by the winding number of the bulk.
Consequently, the total topological charge including contributions
of vortices and edge defects must be zero in a head-to-head domain
wall.

The edge defects in nanomagnets are analogs of the so called boojums
which exist at the surfaces and interfaces of superfluid $^3$He
\cite{Mermin77,Misirpashaev}. In general, ``boojum" refers to a
topological defect that can live only on the surface of an ordered
medium \cite{Volovik}. Boojums were also predicted and observed in
some liquid crystals \cite{Kleman}. An interesting system which is
closely related to our study of flat nanomagnets is the smectic C
films. Conventionally, the ordered state of a liquid crystal is
described by a unit vector $\hat\mathbf n$ pointing along the long
axis of the constituent molecules. Because order parameters
$\hat\mathbf n$ and $-\hat\mathbf n$ are equivalent, it is known
that vortices with half-integer winding numbers are allowed to exist
in the bulk of nematic liquid crystals. On the other hand, for
smectic C liquid crystals, the in-plane ordering of molecular
orientations is described by an additional 2D unit vector $\hat{\bf
c}$ lying in the smectic planes and pointing to the tilt direction
of $\hat\mathbf n$ \cite{deGennes}. Because rotating a tilted
molecule by 180$^\circ$ around the normal of smectic layers does not
return it to its original configuration, this unit vector $\hat{\bf
c}$, like the magnetization $\hat{\bf m}$, is a true vector. As
discussed below, the vector nature of the order parameter $\hat{\bf
m}$ is important to the confinement of halfvortices at the edge.

In this article we review the structure and energetics of elementary
topological defects in nanomagnets. In contrast to the determination
of the configuration of topological defects in superfluids or liquid
crystals where the energy is dominated by short range interactions,
finding solutions of the vector field $\hat{\bf m}({\bf r})$ for
topological defects in nanomagnets is considerably more difficult
due to the nonlocal nature of dipolar interaction. We approached
this problem from two opposite limits dominated by the exchange and
dipolar interactions, respectively; the results are presented in
Sections~\ref{sec-exchange} and \ref{sec-dipolar}. Edge defects of
smectic C films are discussed in Section \ref{sec-smectic}, where we
also point out the similarities and differences of the two models.
We conclude with a summary of our major results in Section
\ref{sec-conclusion}.

\section{Exchange Limit of flat nanomagnets}
\label{sec-exchange}

The magnetic energy of a ferromagnetic nanoparticle
has two major contributions: the exchange energy $A\int|\nabla
\hat\mathbf{m}|^2 \,d^3{r}$ and the dipolar energy $ (\mu_0/2)\int
|\mathbf{H}|^2\, d^3{r}$. The magnetic field ${\bf H}$ is related to
the magnetization through Maxwell's equations, ${\bm \nabla} \times
\mathbf{H} = 0$ and ${\bm \nabla} \cdot (\mathbf{H} + \mathbf{M}) =
0$. Here we disregard the energy of anisotropy, which is negligible
for soft ferromagnets such as permalloy.

Analytical treatment of topological defects is generally impossible
due to the long range nature of dipolar interaction. One usually
minimizes the energy numerically to find stable structures of the
magnetization field. Nevertheless exact solutions are possible in a
thin-film limit \cite{Kurzke04,Kohn04}:  $t \ll w \ll \lambda^2/t
\ll w \log{(w/t)}$ defined for a strip of width $w$ and thickness
$t$. Here $\lambda = \sqrt{A/\mu_0 M^2}$ is a characteristic length
scale of exchange interaction. In this limit the magnetization only
depends on the in-plane coordinates $x$ and $y$, but not on $z$.
More importantly, the magnetic energy becomes a local functional of
magnetization \cite{Kurzke04,Kohn04}:
\begin{equation}
    E[\hat\mathbf{m}(\mathbf{r})]/At = \int_{\Omega}
    |\nabla \hat\mathbf{m}|^2 \, d^2 r + (1/\Lambda)\int_{\partial
    \Omega} (\hat\mathbf{m} \cdot \hat{\bf n})^2 \, d r.
    \label{eq-Kurzke}
\end{equation}
Here $\Omega$ is the two-dimensional region of the film,
$\partial\Omega$ is its line boundary, $\hat{\bf n}\perp \hat{\bm
\tau}$ is unit vector pointing to the outward normal of the
boundary, and $\Lambda = 4\pi\lambda^2 / t \log{(w/t)}$ is an
effective magnetic length in the thin-film geometry.
Eq.~(\ref{eq-Kurzke}) is the familiar XY model \cite{Chaikin} with
anisotropy at the edge resulting from the dipolar interaction.
Denoting $\hat\mathbf m = (\cos\theta, \sin\theta)$, minimization of
the energy $E$ with respect to $\theta$ yields the Laplace equation
$\nabla^2\theta=0$ in the bulk and boundary condition $\hat{\bf
n}\cdot\nabla\theta = \sin2\left(\theta+\theta_e\right)/\Lambda$ at
the edge.

\begin{figure} [b]
\includegraphics[width=0.65\columnwidth]{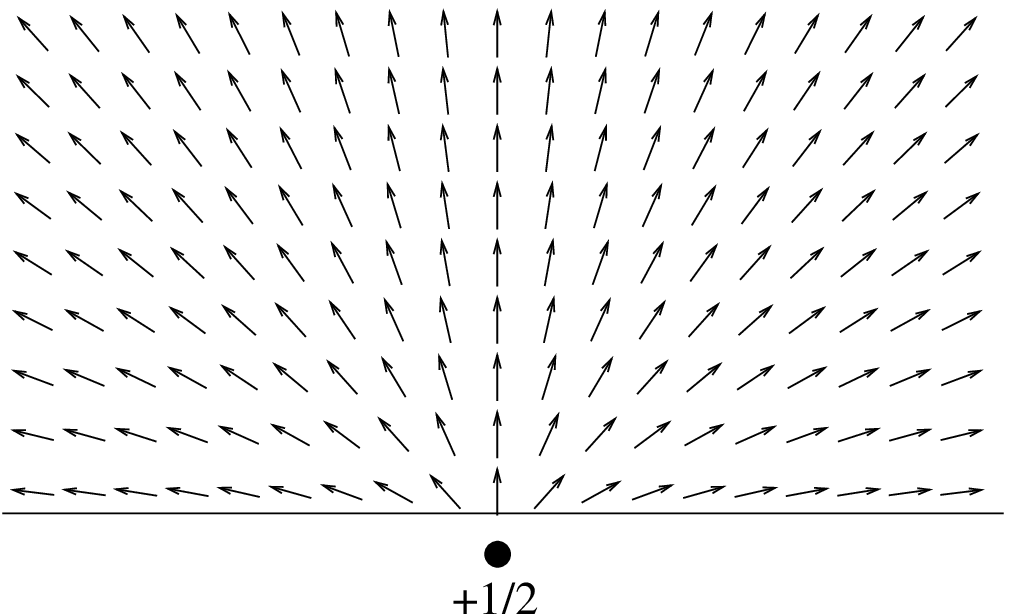}
\includegraphics[width=0.65\columnwidth]{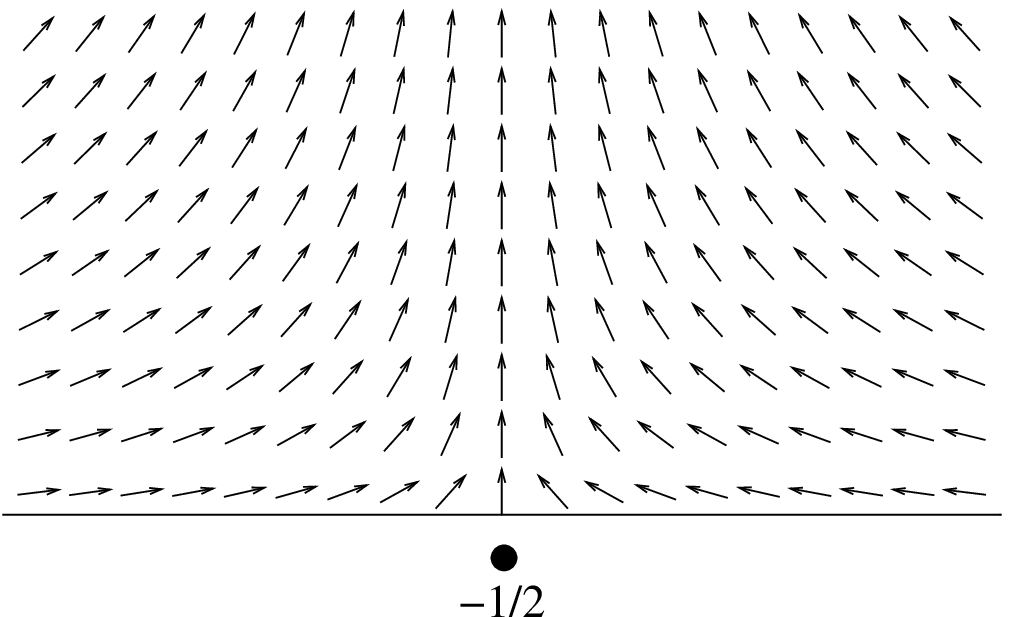}
\caption{Edge defects with winding numbers $n = +\frac{1}{2}$ (top)
and $-\frac{1}{2}$ (bottom) in the exchange limit.}
\label{fig-halfvortex-ex}
\end{figure}

Topological defects that are stable in the bulk are ordinary
vortices with integer winding numbers, which are well known in the
XY model \cite{Chaikin}. The boundary term of model
(\ref{eq-Kurzke}) introduces yet another class of topological
defects that have a singular core outside the edge of the system. To
be explicit, consider an infinite semiplane $y>0$. Solutions
satisfying the Laplace equation in the bulk and the boundary
condition $\partial_y\theta=\sin{2\theta}/\Lambda$ at the edge $y=0$
are \cite{Tch05,Kurzke04}
\begin{equation}
    \tan\theta(x,y) = \pm\; \frac{y+\Lambda}{x-X}.
    \label{eq-halfvortex-ex}
\end{equation}
The singular core is at $(X,-\Lambda)$, distance $\Lambda$ outside
of the edge. Fig.~\ref{fig-halfvortex-ex} shows the magnetization
fields of Eq.~(\ref{eq-halfvortex-ex}). As can be easily checked
using Eq.~(\ref{eq-n-e}) the winding numbers of these solutions are
$\pm\frac{1}{2}$, respectively. The halfvortices can not live in the
bulk: as its singular core is moved inside the boundary, a string of
misaligned spins occurs which extends from the core of halfvortex to
the boundary \cite{Tch05}. The edge thus provides a linear confining
potential for the halfvortices.

In the limit $\Lambda/w\to 0$, magnetization at the edge is forced
to be parallel to the edge, $\hat{\bf m}=\pm\hat{\bm \tau}$. By
exploiting the analogy between XY model and 2D electrostatics, one
can use the method of images to deal with the effects introduced by
the boundary \cite{Chaikin,Tch05}. In this analogy the vortex is
mapped to a point charge whose strength is given by the
corresponding winding number. However, unlike the electrostatics,
the ``image'' charge induced by the boundary has the same sign as
the original. The above solution (\ref{eq-halfvortex-ex}) with
$\Lambda =0$ looks just like a $n=\pm 1$ vortex with its core
sitting at the edge. The assignment of half-integer winding number
$n=\pm\frac{1}{2}$ to the edge defect is thus consistent with the
electrostatics analogy in the sense that the winding number is
doubled by the reflection at the edge \cite{Tch05}.

\begin{figure}
\includegraphics[width=0.75\columnwidth]{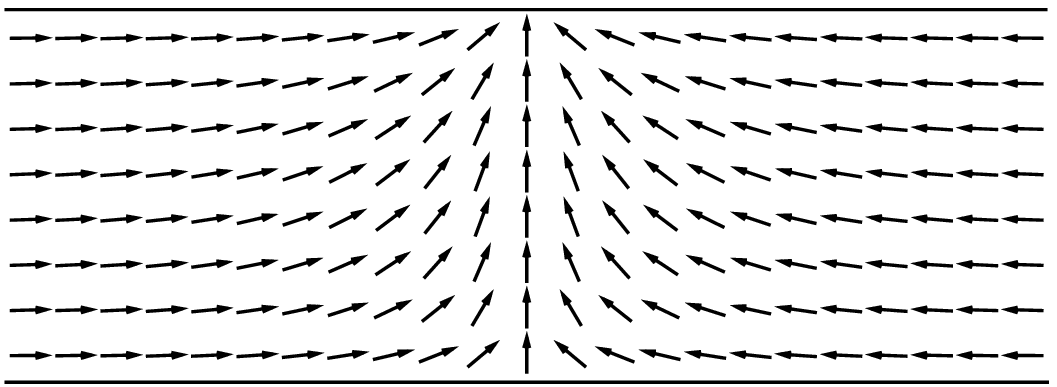}
\includegraphics[width=0.75\columnwidth]{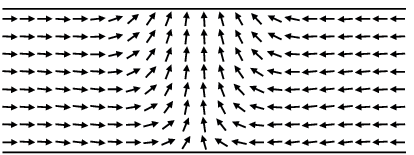}
\caption{Top: Magnetization of the head-to-head domain wall solution
(\ref{eq-trans-wall}). It is composed of two edge defects with
opposite winding numbers $\pm\frac{1}{2}$. Bottom: a transverse
domain wall observed in a micromagnetic simulation using OOMMF
\cite{oommf} in a permalloy strip of width $w= 80$ nm and thickness
$t=20$ nm.} \label{fig-trans-wall}
\end{figure}

An exact solution for a domain wall was also obtained in this limit
\cite{Tch05}. Consider a strip $|y|<w/2$. It has two ground states
with uniform magnetization: $\theta=0$ or $\pi$. Domain walls
interpolating between the two ground states are given by
\begin{equation}
  \tan\theta(x,y) = \pm\frac{\cos{k y}}{\sinh{k (x-X)}},
  \label{eq-trans-wall}
\end{equation}
where the wavenumber $k\approx \pi/(w+2\Lambda)$. The magnetization
field of Eq.~(\ref{eq-trans-wall}) (shown in
Fig.~\ref{fig-trans-wall}) is reminiscent of the so called
`transverse' domain walls (Bottom panel of
Fig.~\ref{fig-trans-wall}) observed in micromagnetic simulations
\cite{McMichael97}.

Unlike domain walls (kinks) in Ising magnet, the domain wall
described by Eq.~(\ref{eq-trans-wall}) is a composite object
containing two edge defects with opposite winding numbers
$\pm\frac{1}{2}$. The singular cores of the two halfvortices reside
outside the film, a distance $\Lambda$ away from the edges. One can
understand the stability of the domain wall using the electrostatics
analogy: the attractive `Coulomb' force pulling together the two
halfvortices is balanced by the confining force from the edges.

The total energy of the domain wall solution
Eq.~(\ref{eq-trans-wall}) evaluates to $E\approx2\pi
At(1+\log(w/\pi\Lambda))$. As expected for the XY model, the
exchange energy depends logarithmically on the system size which is
the width of the strip $w$ in our case. It also depends
logarithmically on a short distance cutoff which is provided by
$\Lambda$ here. After restoring the energy units and expressing
$\Lambda$ in terms of the relevant parameters, we obtain the
following domain wall energy in the exchange limit
\begin{equation}
E_{\rm DW} \approx 2\pi At\log\Bigl(\frac{e w t
\log(w/t)}{\pi\lambda^2}\Bigr). \label{e-dw}
\end{equation}
The energy depends linearly on the thickness of the film $t$ and
only weakly (logarithmically) on the width. These relations are
important to the understanding of the hysteresis curves of
asymmetric magnetic nanorings \cite{FQZhu06}.

\section{Dipolar Limit of flat nanomagnets}
\label{sec-dipolar}

The thin-film limit discussed in the previous section is
inaccessible to most experimental realizations of nanomagnets, in
which the dipolar interaction is the primary driving force. In this
section we discuss the structure and energetics of topological
defects and domain walls in the opposite limit where the energy is
dominated by the dipolar interaction. Our strategy here is first to
find structures which minimize the magnetostatic energy
$(\mu_0/2)\int|{\bf H}|^2\,d^3{r}$ and then to include exchange
interaction as a perturbation. However, energy minimization in the
dipolar limit is relatively difficult due to following reasons.
Firstly, as opposed to the local exchange interaction, the dipolar
interaction is long-ranged. Secondly, in many cases the
magnetostatic energy has a large number of absolute minima. One thus
has to search among these minima for one with the lowest exchange
energy, making it a degenerate perturbation problem.

\begin{figure}
\includegraphics[width=0.6\columnwidth]{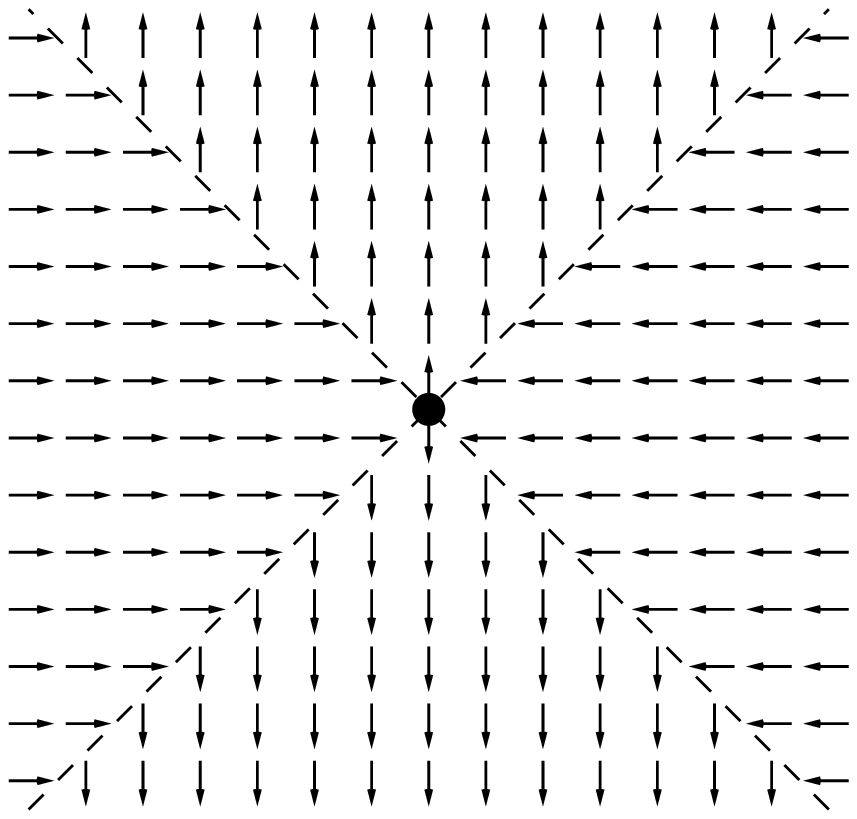}
\includegraphics[width=0.65\columnwidth]{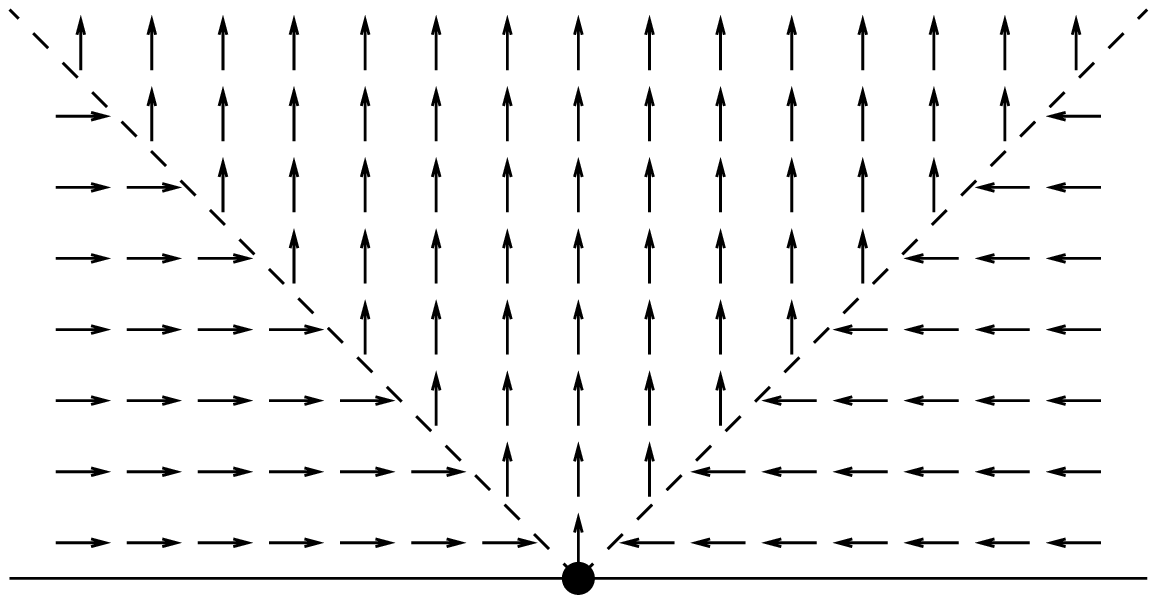}
\caption{an antivortex (top), and a $-\frac{1}{2}$ edge defect
(bottom) in the dipolar limit.} \label{fig-anti-dp}
\end{figure}

The magnetostatic energy of a given magnetization field $\hat{\bf
m}({\bf r})$ can also be expressed as the Coulomb interaction of
magnetic charges with density $\rho_m({\bf r}) =
-M_0\nabla\cdot\hat\mathbf m$, where $M_0$ is the saturation
magnetization. Being positive definite, the magnetostatic energy has
an absolute minimum of zero, which corresponds to a complete absence
of magnetic charges. A general method to obtain the absolute minima
of magnetostatic energy was provided by van den Berg in 1986
\cite{Berg86}. For magnetic films with arbitrary shapes, his method
yields domains of slowly varying magnetization separated by
discontinuous N\'eel walls. In the following we look for structures
that have the desired winding number and are free of magnetic
charges, i.e. $\nabla\cdot\hat{\bf m} = 0$ in the bulk and $\hat{\bf
n}\cdot\hat{\bf m}=0$ on the boundary.

We start by examining the vortex solutions of XY model. In polar
coordinate, a vortex with winding number $n$ is described by
$\theta(x,y) = n\phi+\theta_0$, where $\theta_0$ is a constant
and $\phi=\arctan(y/x)$ is the azimuthal angle. Among these solutions,
only the $n=1$ vortex with $\theta_0=\pi/2$ has zero charge density
and survives in the dipolar
limit. Its energy then comes entirely from the exchange interaction
and diverges logrithmically with system size $R$: $E\approx 2\pi
At\log(R/\lambda)$. Here the short distance cutoff is given by the
exchange length $\lambda$.

The antivortex solutions of the XY model always carry a finite
density of magnetic charge and thus are not a good starting point to
obtain the $n=-1$ defect in the dipolar limit. Fortunately, a
magnetization field with winding number $-1$ and free of bulk
charges is realized by a configuration known as the cross tie
\cite{Chern06,Loehndorf96} (top panel of Fig.~\ref{fig-anti-dp}). It
consists of two $90^\circ$ N\'eel walls normal to each other and
intersecting at the singular core. The magnetization field of an
antihalfvortex (winding number $-\frac{1}{2}$) is obtained by
placing the core of a cross tie at the edge of the film (bottom
panel of Fig. \ref{fig-anti-dp}). Since the magnetization along the
edge is parallel to the boundary, the structure is also free of
surface charge. As one moves from left to right along the edge the
magnetization rotates counterclockwise through $\pi$. This is in
agreement with the definition~(\ref{eq-n-e}) for an antihalfvortex.

\begin{figure}[b]
\includegraphics[width=0.95\columnwidth]{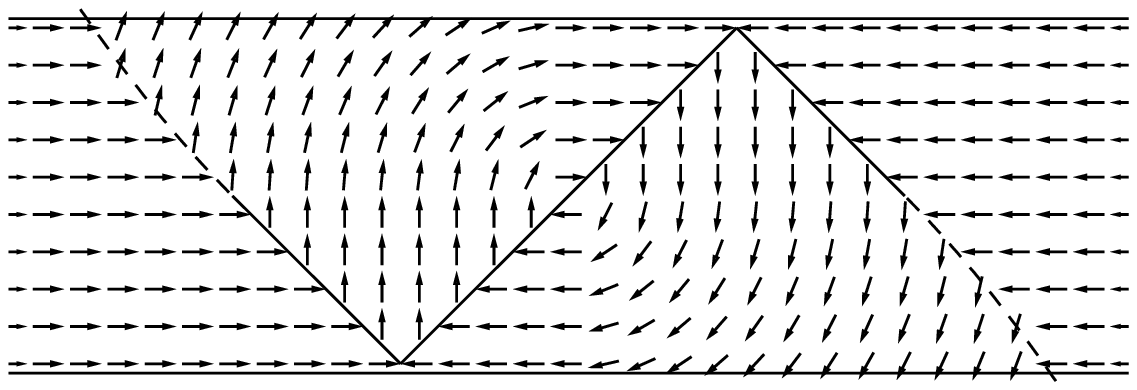}
\includegraphics[width=0.95\columnwidth]{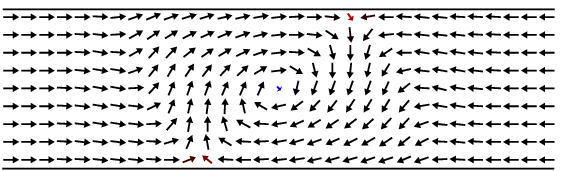}
\caption{Top: a magnetization configuration free of bulk magnetic
charges, $-\nabla \cdot \mathbf{M} = 0$, and containing two $-1/2$
edge defects and a $+1$ vortex in the middle.  Parabolic segments of
Neel walls are shown by dashed lines.  Bottom: a head-to-head vortex
wall obtained in a micromagnetic simulation using OOMMF \cite{oommf}
in a permalloy strip of width $w = 500$ nm and thickness $t = 20$
nm.  } \label{fig-vortex-wall}
\end{figure}

The energy of an antivortex or an antihalfvortex grows linearly with
the length of the N\'eel walls $L$ emanating from it:
\begin{equation}
    E\sim\sigma t L + E_{\rm core}
\end{equation}
The surface tension of the wall $\sigma$ has contributions from both
exchange and dipolar interactions. In magnetic films with thickness
exceeding the N\'eel-wall width (of order $\lambda$), it is given by
\cite{Chern06}
\begin{equation}
    \sigma = 2\sqrt{2}(\sin\theta_0-\theta_0\cos\theta_0)\,A/\lambda,
    \label{eq-wall-tension}
\end{equation}
where $2\theta_0$ is the angle of magnetization rotation across the
wall. In thinner films ($t \lesssim \lambda$) the magnetostatic term
becomes substantially nonlocal and the N\'eel walls acquire long
tails \cite{Hubert}.

There is no charge-free configuration for the $+\frac{1}{2}$ edge
defect. In addition, one could also observe from micromagnetic
simulations that most of the magnetic charges of a transverse domain
wall in a strip is accumulated around the $+\frac{1}{2}$ defect.
Thus, in the dipolar limit, the chargeful $+\frac{1}{2}$ defect is
prone to decay into a $-\frac{1}{2}$ edge defect and $+1$ vortex in
the bulk. We next turn to the discussion of the structure of domain
walls in this limit.

An intrinsic problem arises when one tries to apply van den Berg's
method to find the structure of domain walls. That is because a
head-to-head domain wall carries a fixed nonzero amount of magnetic
charge: $Q_m = 2 M_0 t w$. However, these magnetic charges tend to
repel each other and spread over the surface of the sample, much the
same as the electric charges do in a metal. Based on this principle,
we provided in Ref.~\cite{Youk06} a construction of the head-to-head
domain wall that is free of {\em bulk} magnetic charges. All of the
charge $Q_m$ is expelled to the edges. The resulting structure is
shown in the top panel of Fig.~\ref{fig-vortex-wall}. It resembles
the structure known as the `vortex' domain wall (bottom panel of
Fig. \ref{fig-vortex-wall}) predicted to be stable in regimes
dominated by dipolar interaction \cite{McMichael97}. Both structures
contain two $-\frac{1}{2}$ edge defects sharing one of their N\'eel
walls and a $+1$ vortex residing at the midpoint of the common wall.

The variational construction contains charge-free domains with
uniform and curling magnetization separated by straight and
parabolic N\'eel walls. In a strip $|y|<w/2$, the two $-\frac{1}{2}$
edge defects share a N\'eel wall $x=y$ where the vortex core $(v,v)$
is located. The two curling domains in the regions $\pm v < \pm y
<w/2$ are separated by parabolic N\'eel walls $(x-v)^2 = (2y \pm
w)(2v\pm w)$ from domains with horizontal magnetization; they also
merge seamlessly with other uniform domains along the lines $x = v$
and $y = v$.

\begin{figure} [b]
\includegraphics[width=0.95\columnwidth]{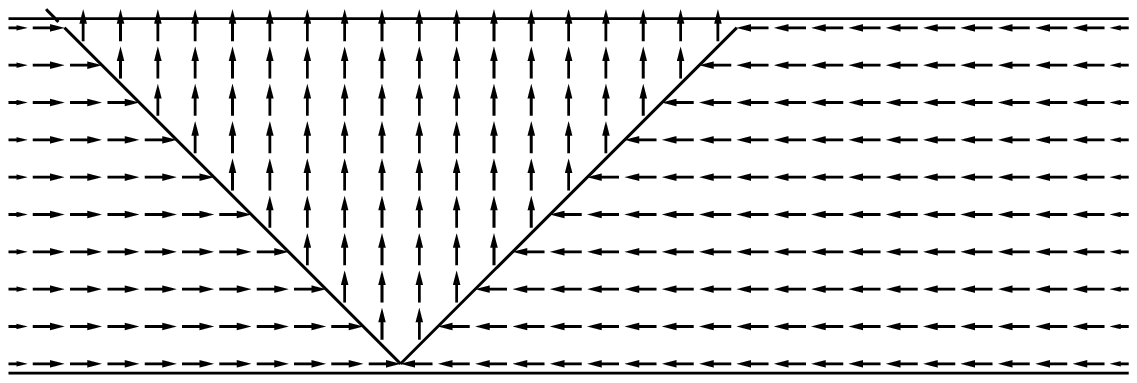}
\includegraphics[angle=-90,width=0.95\columnwidth]{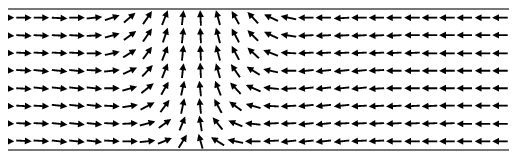}
\caption{Top: A model vortex wall when the vortex is absorbed by
the edge forming an extended $+\frac{1}{2}$ edge defect along
the upper boundary. Bottom: Transverse wall observed in micromagnetic
simulation.}
\label{fig-off-center}
\end{figure}

The location $(v,v)$ of the $+1$ vortex on the shared N\'eel wall is
a free parameter of our variational construction. The structure
remains free of bulk charge as the vortex core moves along the
diagonal $x=y$. When it reaches one of the edge, its annihilation
with the $-\frac{1}{2}$ edge defect creates a widely extended
$+\frac{1}{2}$ edge defect. The resulting structure (top panel of
Fig. \ref{fig-off-center}) is topologically equivalent to the
transverse domain wall (bottom panel of Figs.~\ref{fig-trans-wall}
and \ref{fig-off-center}) discussed in the previous section.

The equilibrium structure for given strip width $w$ and thickness
$t$ is determined by minimizing the total energy of the composite
domain wall with respect to vortex coordinate $v$. The total energy
contains the following terms. (a) The exchange energy of the two
curling domains $\Omega$ around the vortex core. It is given by $A t
\int_\Omega (\nabla\theta)^2 d^2{r}$ and of the order $A t
\log(w/\lambda)$. (b) The energy of N\'eel walls, which can be
computed as a line integral $t\int\sigma(\ell)\,d\ell$, where
$d\ell$ is a line element of the wall. The surface tension $\sigma$,
which depends on the angle of spin rotation across the wall, is
given by Eq.~(\ref{eq-wall-tension}). This term is of order
$Atw/\lambda$. (c) The magnetostatic energy coming from the Coulomb
interaction of magnetic charges spreading along the two edges. It is
of the order $Aw(t^2/\lambda^2)\log(w/t)$.

\begin{figure}
\includegraphics[width=0.95\columnwidth]{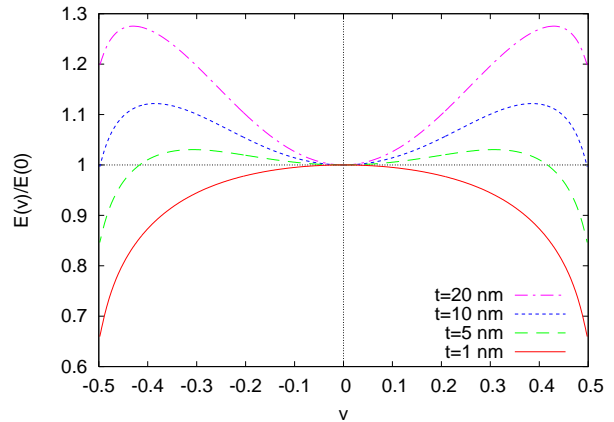}
\caption{Energy of the vortex domain wall as a function of the
vortex position $v$ at a fixed strip width $w = 50$ nm for several
thicknesses $t$.} \label{fig-e-v}
\end{figure}

By combining the above three contributions, the total energy curve
$E(v)$ for a fixed width $w$ and varying thickness $t$ is shown in
Fig.~\ref{fig-e-v}. For substantially wide and thick strips, the
curve attains its absolute minimum as the vortex is in the middle of
the strip, in agreement with numerical simulations
\cite{McMichael97}. A local minimum develops with the vortex core at
the edge of the strip as the thickness decreases. This solution
corresponds to the transverse wall shown in
Fig.~\ref{fig-off-center}. The transverse wall becomes the absolute
minimum as the thickness is further reduced and the vortex wall
($v=0$) is locally unstable. It should be noted that the above
calculation for thin films, e.g. $t=1$ nm, is only an extrapolation.
For films with small cross section (but not in the exchange limit),
our variational approach can not be trusted. Nonetheless, the method
is illustrative and indeed shows that the three-defects wall
structure is unstable when approaching the exchange limit.

\section{Halfvortices in smectic films}
\label{sec-smectic}

The XY model, applicable in the thin-film limit, preserves a
symmetry between topological defects with opposite winding numbers,
namely, $\pm1$ vortices have exactly the same energy in model
(\ref{eq-Kurzke}) (so do $\pm\frac{1}{2}$ edge defects). Since
vortices of opposite winding numbers carry different magnetic
charges, the degeneracy is lifted in thicker and wider strips where
the dipolar interaction becomes more important. The configuration of
the topological defects in the extreme dipolar limit discussed
previously clearly shows this asymmetry. One can also break this
symmetry by assigning different penalties to splay
($\nabla\cdot\hat\mathbf{m}\neq0$) and bending ($\nabla\times\hat
\mathbf{m}\neq0$) deformations:
\begin{eqnarray}
    E[\hat\mathbf{m}(\mathbf{x})] &= &\int_\Omega  \ \left[K_1 (\nabla \cdot
    \hat\mathbf{m})^2 + K_2 (\nabla \times \hat\mathbf{m})^2\right]\,d^2 r
    \nonumber \\
    & & \quad\quad + (1/\Lambda) \int_{\partial \Omega} \ (\hat{\bf n} \cdot
    \hat\mathbf{m})^2\,dr, \label{eq-elastic}
\end{eqnarray}
Here the elastic constants $K_1$ and $K_2$ have energy unit, whereas
the edge anisotropy $1/\Lambda$ scales as the inverse length times
energy. With the unit vector $\hat{\bf m}$ identified as the
$\hat{\bf c}$-director field, this energy functional also describes
the elastic energy of a chiral smectic film \cite{Langer86} or a
Langmuir monolayer \cite{Fischer94} with planar boundary conditions.

The case $K_1 = K_2$ corresponds to the XY model and the exchange
limit discussed in Sec.~\ref{sec-exchange}. By choosing $K_1
> K_2$ we discourage splay, which is similar to a penalty for
magnetic charges in the bulk.  The dipolar limit thus corresponds to
the regime where the bend energy is small compared to those of splay
and edge anisotropy. In what follows we focus on the extreme dipolar
limit $K_2=0$.

First, the $+1$ vortex solution $\theta({\bf r})=\phi+\pi/2$ remains
an energy minimum of model (\ref{eq-elastic}) for arbitrary $K_1$
and $K_2$. The $+1/2$ edge defect in the XY limit,
Eq.~(\ref{eq-halfvortex-ex}) with the `$+$' sign, also is a stable
configuration for arbitrary $K$ and $\Lambda$ except that the
singular core is pushed further outside the boundary, a distance
$(1+\epsilon)\Lambda$ away from the edge. Here $\epsilon =
(K_1-K_2)/(K_1+K_2)$.

\begin{figure}
\includegraphics[width=0.65\columnwidth]{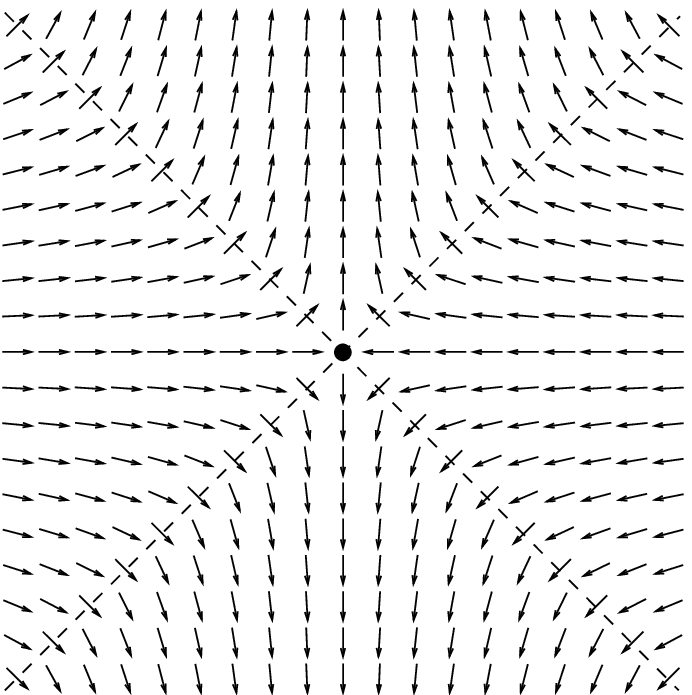}
\includegraphics[width=0.7\columnwidth]{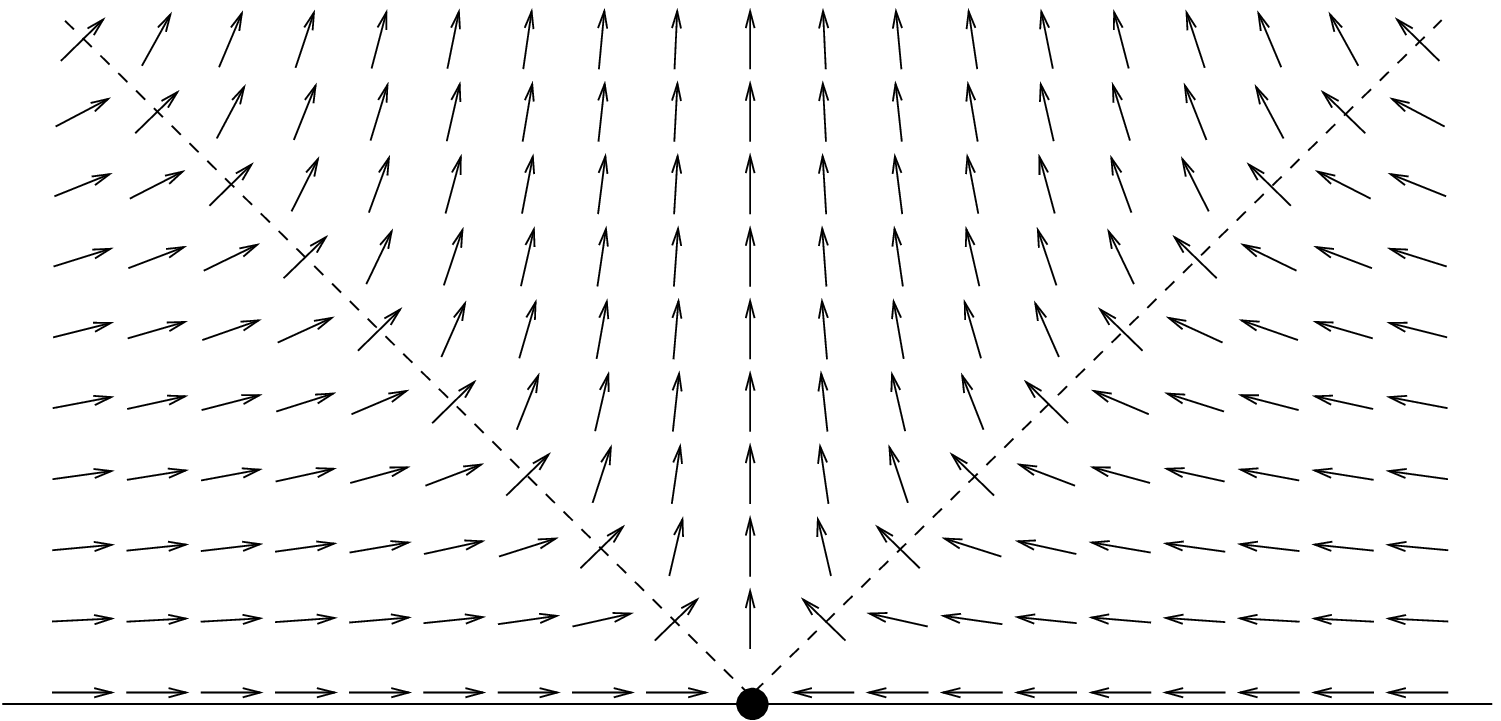}
\caption{an antivortex (top), and a $-\frac{1}{2}$ edge defect
(bottom) in the elastic model with $K_1=1$, $K_2=0$, and $\Lambda =
0$.} \label{fig-anti-smc}
\end{figure}

Since the bulk term of the energy functional (\ref{eq-elastic}) does
not have an intrinsic length scale, an exact scale-invariant
solution for an antivortex has been obtained in the dipolar limit
$K_2=0$:
\begin{equation}
    \theta(x,y)=\phi-\arcsin(\sqrt{2}\,\sin\phi),
    \label{eq-antivortex-sm}
\end{equation}
where $\phi=\arctan(y/x)$ is the azimuthal angle. This solution is
singular at $\phi=\pm\pi/4$, where the first derivative
$d\theta/d\phi$ diverges. A complete solution of the antivortex
nevertheless can be obtained by continuing the above solution
outside of the interval $|\phi|<\pi/4$ periodically. The result is
shown in Fig. \ref{fig-anti-smc}(a).

Analytical solutions of antihalfvortex for arbitrary $\Lambda$ are
yet to be found. In the limit $\Lambda\to 0$ achieved in boundaries
with very strong anchoring force, the unit vector $\hat{\bf m}$ is
forced to be parallel to the edge. In this limit, the $-\frac{1}{2}$
edge defect can be constructed following the same trick for the
antihalfvortex in the dipolar limit of nanomagnets. The resulting
configuration is shown in Fig.~\ref{fig-anti-smc}(b). Compared with
their counterparts in the XY model, the antivortex and
antihalfvortex in Fig.~\ref{fig-anti-smc} are closer to the cross
tie configuration (or half of it) shown in Fig.~\ref{fig-anti-dp}.

Although topological defects of the generalized elastic model show
some similarities with those of the magnetic problem in the dipolar
limit, there is an important difference regarding the scaling of
their energy with system size. Since solution
(\ref{eq-antivortex-sm}) is scale-invariant, the energy of the
defects also diverges logarithmically: $E\sim {\rm const}\times
K_1\log(R/a)$. Here $a$ is a short distance cutoff. In the case of
$-\frac{1}{2}$ edge defect, $a$ is of the order of $\Lambda$.
However, as discussed in the previous section, the energy of both
antivortex and antihalfvortex scales linearly with the length of
N\'eel wall. The nonlocal dipolar interaction in the magnetic
problem results in a natural length scale $\lambda = \sqrt{A/\mu_0
M^2}$. By contrast, there is no such length scale in the elastic
model (\ref{eq-elastic}), so the dependence of the energy on the
system size is logarithmic.

\section{Conclusion}
\label{sec-conclusion}

We have discussed the topological properties of edge defects in XY
systems with a broken O(2) symmetry at the boundary. In particular,
we discussed two physical systems containing such edge defects: the
2D smectic C films and nanomagnets with a planar geometry. Since
spins at the boundary have two degenerate preferred directions, i.e.
parallel or antiparallel to the tangent of boundary, the edge
defects are manifested as kinks of magnetization along the edge.
Moreover, they carry half-integer winding numbers and thus can be
viewed as halfvortices confined to the edge. Conservation of
topological winding number can only be established by including
contributions from the edge defects. As we have pointed out before
\cite{Tch05,Chern06,Youk06}, edge defects should be included along
with ordinary vortices as the elementary topological defects in flat
nanomagnets. Indeed, domain walls which play an significant role in
the dynamics of magnetic nanostrips and nanorings are composite
objects consisting of two or more of these elementary defects.

Analytical solutions of halfvortices and transverse domain walls
were obtained in a thin-film limit where the exchange interaction is
the dominant force determining the shape of topological defects. The
magnetic problem is reduced to the familiar XY model with an
anisotropy at the edge. Domain walls stable in this regime are
composed of two edge defects with winding numbers $\pm\frac{1}{2}$.
By analogy with 2D electrostatics, the stability of transverse
domain wall can be understood as resulting from a balance of the
attractive Coulomb force between the oppositely charged halfvortices
and the confining force from the edges.

Energy minimization is relatively difficult in the opposite limit
dominated by the nonlocal dipolar interaction. Nevertheless, by
focusing on structures which are free of bulk magnetic charge, we
are able to find structures of topological defects stable in this
regime. The $+1$ vortex of XY model with circulating magnetization
remains a stable defect in the dipolar limit. The $-1$ vortex
survives in this limit but is severely deformed; it has the cross
tie structure consisting of two 90$^\circ$ N\'eel walls intersecting
at the singular core. The configuration of the $-\frac{1}{2}$ edge
defect is constructed by placing the core of a cross tie at the
boundary. The $+\frac{1}{2}$ defect carries a finite amount of
magnetic charge and is unstable in this limit.

We have presented a variational construction of the vortex domain
wall which is composed of two $-\frac{1}{2}$ edge defects and a $+1$
vortex. By varying the location of the center $+1$ vortex, the
construction interpolates between the vortex wall and the transverse
wall. Variational calculation of the domain wall energy reveals that
the vortex wall is indeed stable in the dipolar limit whereas it
becomes an energy maximum in thin and narrow strips.

Finally, we have discussed structures of topological defects in an
elastic model which generalizes the XY model of the thin-film limit.
Calculations in this model are simplified by the replacement of
non-local interactions between magnetic charges by a term that
penalizes the existence of magnetic charge in a local fashion. This
model is applicable to smectic C films, but may provide insight into
magnetic configurations. In particular, the allowed topological
defects are the same in both systems. \\

{\bf Acknowledgments}. We thank C.-L. Chien, P. Fendley, D. Huse. R.
L. Leheny, P. Mellado, O. Tretiakov, and F. Q. Zhu for helpful
discussions. The work was supported in part by the NSF Grant No.
DMR05-20491.


\begin{thebibliography}{99}


\bibitem{Chaikin} P. M. Chaikin and T. C. Lubensky, {\em Principles of
Condensed Matter Physics} (Cambridge University Press, Cambridge,
2000).


\bibitem{Mermin79} N. D. Mermin, \rmp {\bf 51}, 591 (1979).


\bibitem{Zhu00} J.-G. Zhu, Y. Zheng, and G. A. Prinz,
J. Appl. Phys. {\bf 87}, 6668 (2000).


\bibitem{Klaeui03} M. Kl\"aui, C. A. F. Vaz, L. Lopez-Diaz and J. A. C. Bland,
J. Phys.: Condens. Matter {\bf 15}, R985 (2003).


\bibitem{Tch05} O. Tchernyshyov and G.-W. Chern,
Phys. Rev. Lett. {\bf 95}, 197204 (2005).


\bibitem{Chern06} G.-W. Chern, H. Youk, and O. Tchernyshyov,
J. Appl. Phys. {\bf 99}, 08Q505 (2006).


\bibitem{Youk06} H. Youk, G.-W. Chern, K. Merit, B. Oppenheimer, and
O. Tchernyshyov, J. Appl. Phys. {\bf 99}, 08B101 (2006).


\bibitem{Mermin77} N. D. Mermin, pp. 3-22 in {\em Quantum Fluids and
Solids}, eds. S. B Trickey, E. D. Adams and J. W. Dufty, (Plenum,
New York, 1977).


\bibitem{Misirpashaev} T. Sh. Misirpashaev, Sov. Phys. JETP {\bf 72}, 973(1991).


\bibitem{Volovik} G. E. Volovik,  {\em The Universe in a Helium Droplet}
(Clarendon Press, Oxford, 2003).


\bibitem{Kleman} M. Kleman and O. D. Lavrentovich, {\em Soft
Matter Physics} (Springer-Verlag, New York, 2003).


\bibitem{deGennes} P. G. deGennes and J. Prost,
{\em The Physics of Liquid Crystals} (Clarendon Press, Oxford,
1993).


\bibitem{Kurzke04} M. Kurzke,
Calc. Var. PDE {\bf 26}, 1 (2006).


\bibitem{Kohn04} R. V. Kohn and V. V. Slastikov,
Proc. Roy. Soc.  (London) Ser. A {\bf 461}, 143 (2005).


\bibitem{McMichael97} R. D. McMichael and M. J. Donahue,
IEEE Trans. Magn. {\bf 33}, 4167 (1997).


\bibitem{FQZhu06} F. Q. Zhu, G.-W. Chern, O. Tchernyshyov,
X. C. Zhu, J. G. Zhu, and C. L. Chien, Phys. Rev. Lett. {\bf 96},
027205 (2006).


\bibitem{Berg86}  H. A. M. van den Berg, J. Appl. Phys. {\bf 60}, 1104
(1986).


\bibitem{Loehndorf96}  M. Londorf, A. Wadas, H. A. M. van den Berg,
and R. Wiesendanger, Appl. Phys. Lett. {\bf 68}, 3635 (1996).


\bibitem{Hubert}  A. Hubert and R. Schaefer, {\em Magnetic Domains}
(Springer, Berlin, 1998).


\bibitem{oommf} M. J. Donahue and D. G. Porter, OOMMF User's Guide,
Version 1.0, in  {\em Interagency Report NISTIR 6376} (NIST,
Gaithersburg, 1999).  http://math.nist.gov/oommf/


\bibitem{Langer86} S. A. Langer and J. P. Sethna,
Phys. Rev. A {\bf 34}, 5035 (1986).


\bibitem{Fischer94} T. M. Fischer, R. F. Bruinsma, and C. M. Knobler,
Phys. Rev. E {\bf 50}, 413 (1994).



\end{thebibliography}
\end{document}